**Measurement of the thermal conductance of silicon nanowires at low temperature**

Olivier Bourgeois[a)], Thierry Fournier and Jacques Chaussy

*Centre de Recherches sur les Très Basses Températures, CNRS, laboratoire associé à l'Université Joseph Fourier et à l'Institut Polytechnique de Grenoble, BP 166, 38042 Grenoble Cedex 9, France.*

Abstract

We have performed thermal conductance measurements on individual single crystalline silicon suspended nanowires. The nanowires (130 nm thick and 200 nm wide) are fabricated by e-beam lithography and suspended between two separated pads on Silicon On Insulator (SOI) substrate. We measure the thermal conductance of the phonon wave guide by the 3ω method. The cross-section of the nanowire approaches the dominant phonon wave length in silicon which is of the order of 100 nm at 1K. Above 1.3K the conductance behaves as $T^3$, but a deviation is measured at the lowest temperature which can be attributed to the reduced geometry.

---

[a)] Electronic mail : olivier.bourgeois@grenoble.cnrs.fr

Thanks to the recent progress in nanofabrication technology, it becomes possible to study the thermal properties of mechanically suspended nanostructures.[1,2] The continuous reduction of size in today's electronic brings new questions. Especially, all the aspects linked to the heat transfer and thermalization at the nanometer scale have to be understood and controlled[3-5]. Particularly, the thermal and thermodynamic behaviours of nano-objects at low temperature remain largely unexplored[6,7]. Some examples of experimental interest are: limitation of thermal transport,[1,8-10] new phase transitions[11], phonon localization[12], entropy or information flow[13-15] etc…The great difficulties linked to the study of thermal properties of nanomaterials lie in the fact that heat flow and temperature measurement have to be controlled at a very small length scale, which makes these experiments usually very challenging. More specifically, while the thermal properties of silicon nitride have been widely studied[16,17], thermal properties at low temperature of nanostructured single-crystalline silicon wire or membrane, widely used in Micro or Nano Electro Mechanical Systems (MEMS or NEMS), have not been published yet.

In this letter, we report thermal conductance measurements performed on individual single-crystalline silicon nanowires mechanically suspended, having a large aspect ratio. The thermal conductance is measured by the 3ω method.[18,19] The nanowires have a rectangular cross-section of the order of the dominant phonon wave length around 1K. We show, through the measurement of the thermal conductance of the nanowire, that the phonon transport in silicon at low temperature follows a regular $T^3$ law. A deviation is however observed at the lowest temperature which might be attributed to the reduced geometry down to the nanometer scale.

The suspended wires are fabricated in the CEA-LETI facilities from Silicon On Insulator (SOI)[20] 200 mm substrate (see the schematic given in Fig. 1). A first step of thermal oxidation on buffered HF etching thins the upper layer down to 130 nm. The wire and the pads are then written on negative resist (Shipley UVN2) using e-beam lithography. The pads and the wire are structured in the upper silicon layer by Reactive Ion Etching down to the SiO$_2$ layer. The SiO$_2$ layer is then removed with buffered HF. The under-etching liberates the Si wire which remains suspended between the two pads. The pads are electrically isolated from the crystalline silicon underneath thanks to the SiO$_2$ layer. Si nanowires with large aspect ratio were obtained with lengths of 5 to 15 μm and cross sections of 130 by 200 nm.

Following these fabrication steps, a transducer is deposited onto the surface of the pads and the wire. A niobium nitride (NbN) thin films transducer is directly AC sputtered onto the surface of the device (pads and nanowire, see Fig. 1). This thermometric element undergoes a metal to insulator transition at low temperature, and hence exhibits a large temperature coefficient of resistance $\alpha = \frac{1}{R}\frac{dR}{dT}$, on the order of $\alpha$=-0.3K$^{-1}$ at 1K.[21]

Our measurement of thermal conductance is based on the 3ω method, explained in detail by Lu et al.[19] This technique is particularly suited for suspended filament-like structures like ours and allows the direct measurement of low thermal energy exchanges in individual nanowires.

The method consists in applying an ac current $I_{ac}=I_o\ sin(\omega t)$ to the transducer deposited on the surface of the suspended wire at a the frequency $\omega$. The power dissipated in the transducer heats both it and the silicon wire, generating an oscillation of temperature at 2$\omega$. If the oscillation of temperature is sufficiently small, the voltage ($V= I_{ac} R(T)$) has two major components, one at $\omega$ and the other one at 3$\omega$. (see reference 19). By solving the differential equation of the thermal balance of this suspended system, Lu, et al. demonstrated that the third harmonic of the voltage (denoted by $V_{3\omega}$) is directly connected to the thermal properties of the nanowire which are the thermal conductance, the heat capacity and the power dissipated through the formula:[19]

$$V_{3\omega} = \frac{4R^2 I_{ac}^3 \alpha}{\pi^4 K \sqrt{1+(2\omega\gamma)^2}} \qquad (1)$$

where $K$ is the thermal conductance and $\gamma$ is the characteristic thermal relaxation time of the system to the heat bath. At high frequency ($\omega > 1/\gamma$) the system can be considered as in the adiabatic limit, $V_{3\omega}$ is dependent only on the specific heat. On the other hand, at low frequency ($\omega < 1/\gamma$) the system is in the quasi-static regime and hence the $V_{3\omega}$ voltage is related to the thermal conductance.

The low frequency limit can be established through the measurement of the frequency dependence of $V_{3\omega}$. As shown in Fig.2, the voltage measured across the transducer at $3\omega$ decreases as the measurement frequency is increased following the equation 1. From a fit of the experimental data using the equation 1, the thermal characteristic time of the systems is extracted at 1 K and found to be of the order of 4 ms. By working around 4 Hz, we can consider that $\omega\gamma \ll 1$, the signal is then only dependent on the thermal conductance of the wire. We now consider that the thermal conductance is correctly given by the formula:[19]

$$K = \frac{4R^2 I_{ac}^3 \alpha}{\pi^4 V_{3\omega}} \qquad (2)$$

This model applies only in the case of wires having highly rough surface which might not be entirely the case in our systems; as demonstrated by Holmes *et al.* this strongly depends on the surface state of the phonon wave guide.[17]

The system is actually constituted by the silicon nanowire and on top of it the transducer. The silicon wire has to be indeed the principal thermal conductor. Therefore, one has to make sure that no major heat loss appears through the transducer as a parallel thermal path. The geometry of the wire imposes an electrical resistance of the tranducer varying from 50 to 100 kΩ between 0.5 and 4 K. We then evaluate the expected thermal conductance of the transducer through the Wiedemann-Franz law valid for amorphous materials at low temperature,[22] where the thermal conductance is dominated by the transport of heat by the electrons. Using this relation, we calculated the thermal conductance expected for the transducer on the order of $10^{-13}$ W/K in the better case, which is two orders of magnitude smaller than that expected for silicon ($10^{-11}$ W/K at 1K) over the entire temperature range of our measurements.

The voltage appearing across the transducer is measured via a traditional four probe technique (connected on the pads of the wires) in a cryostat down to 0.5K. The measurements is performed at 4.1 Hz (frequency of the electrical current applied to the transducer) on wires having different length ranging from 5 to 10µm with section of 130x200nm$^2$. At 0.6 K, the measured rms voltage noise is of the order of 4 nV/√Hz as compared to the signal at $3f$ of the order of 400 nV. An averaging over 1 to 2 minutes leads to a resolution of $\Delta K/K=10^{-3}$ giving a sensitivity of $7\times10^{-15}$ W/K (i.e. $10^{-5}$ W/cmK). The typical amplitude of the oscillation of temperature of the central point of the wire is given by the relation: $\Delta T_{ac} = \dfrac{\pi V_{3\omega}}{4 R I_{ac} \alpha}$. The amplitude of the ac current is chosen to give an oscillation of 1mK. This device can then detect power variations below $10^{-17}$ W/√Hz.

The temperature variation of the thermal conductance of a 5 µm long nanowire is shown on the Fig. 3. On the inset of the Fig. 3, $K$(W/K) is plotted versus $T^3$ to highlight the regular cubic power law behaviour of the thermal conductance. At low temperature the phonon transport in single crystalline silicon is ballistic; the phonon mean free path is only limited by the geometry of the wire. Hence the phonon transport is governed by the boundary scattering on the rough surfaces of the wires; this limit is known as the Casimir regime.[23-25] The expected thermal conductance for a long perfect crystal with rough surfaces can be estimated in the Casimir model by [23-25]:

$$K = \frac{2\pi^2 k_B^4}{15 \hbar^3} \frac{\ell S}{v_s L} T^3 \qquad (3)$$

where $\ell$ is the phonon mean free path, $v_s$ the speed of sound in silicon ($v_s$= 4500m/s)[26,27], $S$ the surface of the wire section, $L$ the relevant length of the wire; in our case $L$=2.5μm because in the 3ω method it can be considered that the heat is deposited in the middle of the wire and flowed to the heat bath through two thermal paths. The fit performed on the data shown in Fig. 3 gives a variation versus temperature of the thermal conductance of $K(T) = 2.6 \times 10^{-11} T^3$ W/K, when the only fitting parameter, the phonon mean free path $\ell$, is set to 0.62 μm. This value is larger than the dimension of the section of the wire indicating that specular reflections on the edge occur during the phonon transport and play a non negligible role.

At lower temperature, a clear deviation from the cubic power law is observed, as it is shown on the Fig. 3. This can be explained by the increase of the mean free path as the temperature is lowered. Indeed, at low temperature, the thermal conductance takes the form $K = \frac{1}{3} C v_s \ell$ where $v_s$ is the average sound velocity, $C$ the specific heat of the lattice (proportional to $T^3$) and $\ell$ the mean free path. By decreasing the temperature the dominant phonon wave length $\lambda_D$ (given by $\lambda_D = h v_s/(2.82 k_B T)$, where $h$ the Planck constant)[23] becomes of the order of the characteristic length of the asperity $\eta$ of the nanowire. As explain by Ziman in reference 23, this plays a major role in the scattering process. When $\lambda_D$ becomes much larger than $\eta$, the specular reflections of phonon on the surface are restored as the temperature is decreased and hence the mean free path increases. In silicon, $\lambda_D$ is of the order of 100nm at 1K, the same order of magnitude that the section of the wires, this can be estimated to be ten times bigger than the asperity (1 to 5nm) of the vertical edges of the nanowire. So the deviation from the $T^3$ law observed at the lowest temperature might be explained by the expected significant increase of the phonon mean free path in the small wires.

To conclude, we have measured the thermal conductance of nanostructured Si wires at low temperature. We have shown that the 3ω method is well suited for the measurement of small thermal signal on silicon nanowires. Accurate value for the thermal conductance is found for silicon at low temperature which can be of great importance for the future design of suspended membrane for calorimetric measurements or for the conception of bolometric devices, MEMS or NEMS.

The authors want to thank E. André, P. Lachkar, and J-L. Garden for technical support, the CEA/LETI/PLATO for the realization of the silicon wires and also P. Gandit, P. Brosse-Maron, J-L. Bret and the electronic shop for useful help, assistance and discussions.

Figure 1. Up: Scanning Electron Microscope image of a suspended silicon nanowire (dimensions: 5 μm long, 130 nm thick and 200 nm wide). The schematic of the device is given on the bottom of the figure showing the suspended silicon wire with the thin film thermometer (100 nm thick). The pads are constituted by an electrical insulating oxide layer (the 400 nm thick $SiO_2$ layer) and the Si upper layer.

Figure 2. Frequency dependence of the 3$\omega$ voltage across the transducer obtained at 1 K. The solid line is a fit according to the equation (1) (see text). The correct working frequency range given by the low frequency limit of the signal lies within few percent of the maximum voltage: between 1 and 5 Hz. This theoretical fit gives the characteristic time scale of the heat loss to the thermal bath: $\gamma$=4 ms.

Figure 3. Thermal conductance measurement made at 4.1 Hz on a 5 μm long silicon nanowire versus the temperature, at the lowest temperature a deviation from the $T^3$ law is observed. Inset: the thermal conductance follows a regular $T^3$ law above 1.4 K and shows a deviation at low temperature. The solid line is a fit of the linear part of the curve (see text).

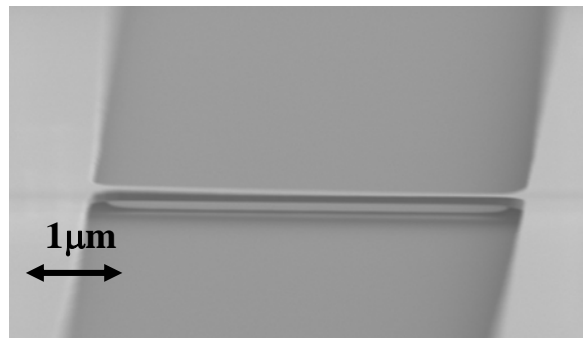

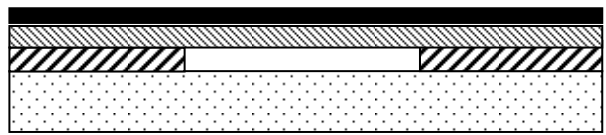

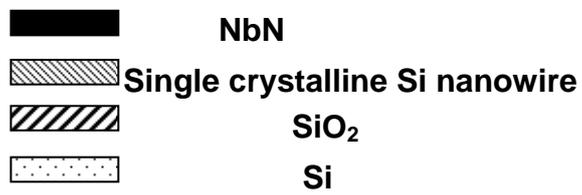

Figure 1, O. Bourgeois *et al*.

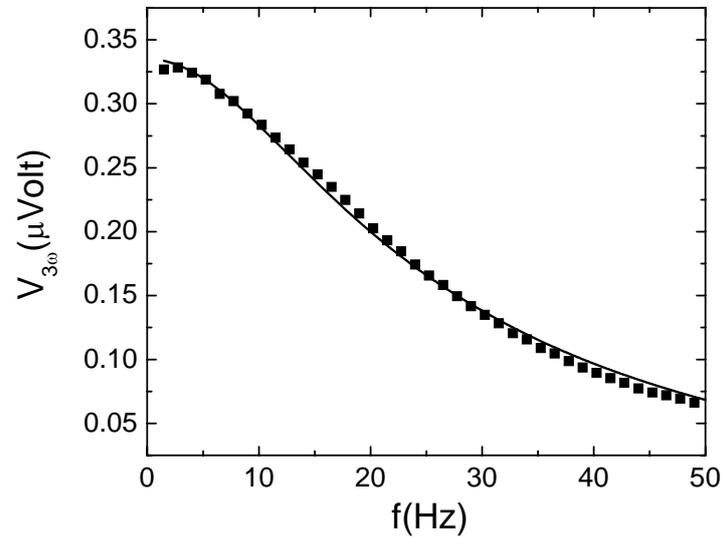

Figure 2, O. Bourgeois *et al*.

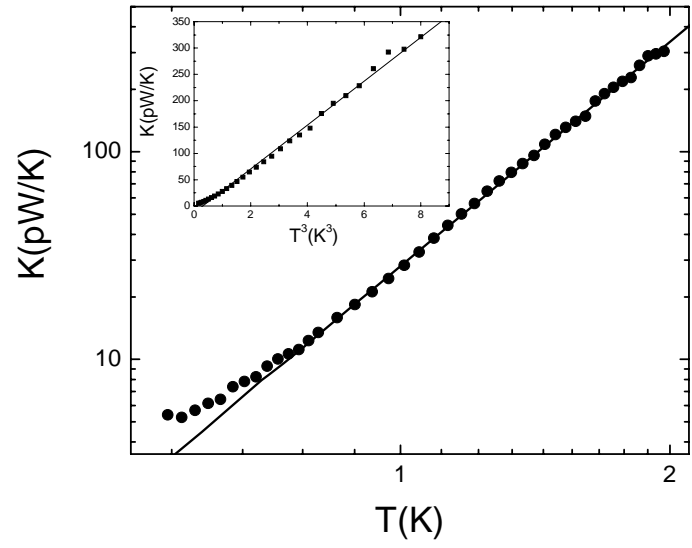

Figure 3, O. Bourgeois *et al*.